\def\be{\begin{equation}}
\def\ee{\end{equation}}
\def\ba{\begin{array}}
\def\ea{\end{array}}

\def\Cb{{\Bbb C}}

\documentclass[preprint,showpacs,amsmath]{revtex4}
\usepackage{amssymb}
\def\qed{\leavevmode\unskip\penalty9999 \hbox{}\nobreak\hfill
     \quad\hbox{\leavevmode  \hbox to.77778em{%
               \hfil\vrule   \vbox to.675em%
               {\hrule width.6em\vfil\hrule}\vrule\hfil}}
     \par\vskip3pt}

\newtheorem{theorem}{Theorem}
\newtheorem{corollary}{Corollary}

\input amssym.def
\begin{document}

\begin{center}
A Note on Resistance of NPT to Mixture of Separable States
\end{center}

\begin{center}
Bobo Hua$^{1}$, Xiu-Hong Gao$^{2}$, Ming-Jing Zhao$^{1}$ and Shao-Ming Fei$^{1,2}$

\vspace{2ex}

{\small $~^{1}$ Max-Planck-Institute for Mathematics in the Sciences, 04103
Leipzig, Germany}

{\small $~^{2}$ School of Mathematical Sciences, Capital Normal
University, Beijing 100048, China}

\medskip
\end{center}

{\bf Abstract}. We study the stability of NPT property of an
arbitrary pure entangled state under the mixture of arbitrary pure separable
states. For bipartite pure states with Schmidt number $n$
$(n>1)$ which is NPT, we show that this
state is still NPT when it is mixed with no more than $\frac{n(n-1)}{2}-1$ arbitrary
pure separable states. This result is generalized to
multipartite case.

{\it Keywords}: partial transposition; separable state; entanglement

{PACS Numbers: 03.65.Ud, 03.67.Mn}

\bigskip
\bigskip

Quantum entangled states are used as key resources in quantum information processing
such as quantum teleportation, cryptography, dense coding, error correction
and parallel computation.
Due to the decoherence maximally entangled pure states may evolve into non-maximally
entangled ones. Distillation is an important protocol in improving the
quantum entanglement against the decoherence due to noisy
channels in information processing.
However, there are two kinds of quantum entangled states, distillable and
non distillable.

Let $H$ be a $d$-dimensional complex Hilbert spaces, with $\{\vert i\rangle\}_{i=1}^d$
the orthonormal basis of the space $H$.
Any bipartite quantum state $\rho\in H\otimes H$ can be written as
$\rho=\sum_{i,j,k,l} \rho_{ij,kl} |ij\rangle\langle kl|$, $\rho_{ij,kl}\in \Cb$.
The partial transposition of $\rho$ with respect to the first (resp. second) system is
$\rho^{T_1}=\sum_{i,j,k,l} \rho_{ij,kl} |kj\rangle\langle il|$ (resp.
$\rho^{T_2}=\sum_{i,j,k,l} \rho_{ij,kl} |il\rangle\langle kj|$).
The transpositions with respect to the two systems
are related by $\rho^{T_1}=(\rho^{T_2})^T$, with $T$
denoting the transposition of the whole matrix. Hence the positivity of
$\rho^{T_1}$ is equivalent to the positivity of $\rho^{T_2}$.
A quantum state that its partially transposed matrices $\rho^{T_1}$ and $\rho^{T_2}$ are positive
is called a PPT (positive partial transposition) state.
It has been shown \cite{8} that any entangled PPT states are not distillable.
These states are called bound entangled \cite{UPB,P. Horodecki1999,9,piani,fei-li-sun,7}.

If $\rho^{T_1}$ and $\rho^{T_2}$ have negative eigenvalues, the state $\rho$ is called NPT (non positive partial transposition).
For example, all entangled pure states, entangled isotropic states \cite{M.
Horodecki1999} and entangled Werner states \cite{werner} are all NPT.
Moreover, an NPT state is necessarily entangled \cite{A. Peres} and believed to be free entangled (distillable).
NPT states are significant resources for quantum information and quantum computation \cite{R. Horodecki}.

In this paper, we study the stability of NPT property of an
arbitrary pure entangled state, namely the resistance of an entangled pure state
to the mixture with pure separable perturbation.
Suppose $|\chi_0\rangle$ is any bipartite entangled pure state.
Let $|\chi_i\rangle$, $i=1,...,K$, be arbitrary pure separable states.
Consider the mixed quantum state,
\begin{eqnarray}\label{main state}
\rho=\lambda_0 |\chi_0\rangle\langle\chi_0| + \sum_{i=1}^{K} \lambda_i |\chi_i\rangle\langle\chi_i|,
\end{eqnarray}
where $0<\lambda_i<1$, $i=0,1,\cdots, K$, $\sum_{i=0}^{K}\lambda_i=1$.
It is interesting to ask how large the number $K$ can be such that $\rho$ is still NPT.
For bipartite pure state $|\chi_0\rangle$ with Schmidt number $n$
$(n>1)$, we show that $\rho$ is still NPT for
$K\leq \frac{n(n-1)}{2}-1$. This result is then generalized to multipartite case.

Denote $n$ $(n>1)$ the Schmidt number of the state $|\chi_0\rangle$.
Under some local unitary transformations $|\chi_0\rangle$ can be expressed in Schmidt form,
$|\chi_0\rangle=\sum_{i=1}^{n} \mu_i|ii\rangle$ , $\mu_i>0$, $\sum_{i=1}^{n} \mu_i^2 =1$.
The state $|\chi_0\rangle\langle\chi_0|$ is NPT, because the eigenvalues of $(|\chi_0\rangle\langle\chi_0|)^{T_1}$ are $\mu_i^2$, $\pm \mu_i \mu_j$, $i,j=1,2,\cdots, n$ and $i\neq j$. Hence
$(|\chi_0\rangle\langle\chi_0|)^{T_1}$ has $\frac{n(n-1)}{2}$
negative eigenvalues. We first present a result for a simple case.

\begin{theorem}\label{th n=2}
If $n=2$ and $K=1$, then the state (\ref{main state}) is NPT.
\end{theorem}

Proof. In this case the state (\ref{main state})
has the form, $\rho=\lambda_0 |\chi_0\rangle\langle\chi_0| + \lambda_1 |\chi_1\rangle\langle\chi_1|$. 
Since $|\chi_0\rangle$'s Schmidt number is 2, there are
exist unitary operators $U$ and $V$  such that
$|\tilde{\chi}_0\rangle\equiv U\otimes V |\chi_0\rangle=\mu_1|11\rangle +\mu_2|22\rangle$. 
Instead of $\rho$, we consider the state 
$\tilde{\rho}\equiv (P\otimes P)(U\otimes V) \rho (U^\dagger\otimes V^\dagger) (P\otimes P)$, 
where $P=|1\rangle\langle1| +|2\rangle\langle2|$ is a project operator.
Then $\tilde{\rho}$ has the following form,
\begin{equation}
\tilde{\rho}=\lambda_0 |\tilde{\chi}_0\rangle \langle \tilde{\chi}_0|+ \lambda_1 |\tilde{\chi}_1\rangle\langle\tilde{\chi}_1|,
\end{equation}
where $|\tilde{\chi}_1\rangle=(P \otimes P)\,(U\otimes V) |\chi_1\rangle$ is still a separable
state, as $|{\chi}_1\rangle$ is separable. Therefore $|\tilde{\chi}_1\rangle$ is
generally of the form,
$|\tilde{\chi}_1\rangle=(a|1\rangle+b|2\rangle)\otimes(c|1\rangle+d|2\rangle)$
with $|a|^2+|b|^2=|c|^2+|d|^2=1$. And the determinant of $\tilde{\rho}^{T_1}$ is given by
$$
-\lambda_0^4\mu_1^4\mu_2^4-\lambda_1\lambda_0^3\mu_1^2\mu_2^2|\mu_1bd+\mu_2ac|^2<0.
$$
Hence $\tilde{\rho}$ is NPT, which implies that the state $\rho$ is NPT too. \qed

\begin{theorem}\label{th1}
If $n>2$ and $K\leq\frac{n(n-1)}{2}-1$, the quantum state (\ref{main state}) is still NPT.
\end{theorem}

Proof. Since $(|\chi_0\rangle\langle\chi_0|)^{T_1}$ has $\frac{n(n-1)}{2}$
negative eigenvalues, the linear subspace $V_{-}$ spanned by all the
eigenvectors associated with the negative eigenvalues of
$(|\chi_0\rangle\langle\chi_0|)^{T_1}$ is $\frac{n(n-1)}{2}$
dimensional, $dim V_{-}=\frac{n(n-1)}{2}$. Note that the dimension of the subspace
$V_{s}$ spanned by $K$ arbitrary separable pure states
$\{|\chi_i\rangle\}_{i=1}^K$ is at most $\frac{n(n-1)}{2}-1$, $dim
V_{s} \leq \frac{n(n-1)}{2}-1$. As the dimension of the whole vector space $H\otimes H$ is $d$,
$d\geq n^2$, the dimension of the
complementary space $V_{s}^{\bot}$ of $V_{s}$ satisfies $dim V_{s}^{\bot}
\geq d+1 -\frac{n(n-1)}{2}$. Subsequently, $dim V_{-} + dim
V_{s}^{\bot}\geq d+1$, which implies that there exists a vector
$|\xi\rangle$ belonging to the intersection of the subspaces $V_{-}$
and $V_{s}^{\bot}$, such that
\begin{eqnarray*}
\langle \xi (|\chi_i\rangle\langle\chi_i|)^{T_1} \xi\rangle&=&\langle \xi |\chi_i\rangle\langle\chi_i| \xi\rangle=0, \ \ \forall i,\\
\langle
\xi|(|\chi_0\rangle\langle\chi_0|)^{T_1} |\xi\rangle&<&0,
\end{eqnarray*}
and further
\begin{eqnarray*}
\langle \xi |\rho^{T_1} |\xi\rangle =\lambda_0 \langle
\xi|(|\chi_0\rangle\langle\chi_0|)^{T_1} |\xi\rangle+ \sum_{i=1}^{K}
\lambda_i  \langle \xi |\chi_i\rangle\langle\chi_i| \xi\rangle<0.
\end{eqnarray*}
Therefore $\rho$ is NPT.
\qed

{\it Comment}. If $K>\frac{n(n-1)}{2}$, then the quantum state $\rho$ in
(\ref{main state}) can be either  NPT, or PPT entangled or PPT separable.
This can be seen from the following examples.

Example 1. Consider the $3\otimes 3$ pure state
$|\chi_0\rangle=0.5|11\rangle+0.8|22\rangle+\sqrt{0.11}|33\rangle$, and the following four
separable states,
\begin{eqnarray}
|\chi_1\rangle&=&(0.4|1\rangle-0.6|2\rangle+\sqrt{0.48}|3\rangle) \otimes (0.3|1\rangle+0.95|2\rangle+\sqrt{0.0075}|3\rangle),
\nonumber\\[2mm]
|\chi_2\rangle&=&(0.27|1\rangle+0.5|2\rangle+\sqrt{0.6771}|3\rangle) \otimes (-0.75|1\rangle-0.1|2\rangle+\sqrt{0.4275}|3\rangle),
\nonumber\\[2mm]
|\chi_3\rangle&=&(-0.2|1\rangle+0.4|2\rangle+\sqrt{0.8}|3\rangle) \otimes (-0.05|1\rangle+0.01|2\rangle-\sqrt{0.9974}|3\rangle),
\nonumber\\[2mm]
|\chi_4\rangle&=&(0.2|1\rangle+0.6|2\rangle-\sqrt{0.6}|3\rangle)  \otimes (0.8|1\rangle-0.55|2\rangle-\sqrt{0.0575}|3\rangle).
\nonumber
\end{eqnarray}
Take $\lambda_0=0.01, \lambda_1=0.6,
\lambda_2=0.09,\lambda_3=\lambda_4=0.15$. In this case $n=3$, $K=4$. The quantum state $\rho$ in
Eq. (\ref{main state}) is PPT because the minimal eigenvalue of
$\rho^{T_1}$ is $0.00006$.

Example 2. The Horodecki's $3\otimes 3$ state \cite{P. Horodecki1999},
\begin{eqnarray*}
\sigma_{\alpha}=\frac{2}{7}|\psi^+\rangle\langle \psi^+|
+\frac{\alpha}{21}
(|01\rangle\langle01|+|12\rangle\langle12|+|20\rangle\langle20|) +
\frac{5-\alpha}{21}
(|10\rangle\langle10|+|21\rangle\langle21|+|02\rangle\langle02|),
\end{eqnarray*}
where
$|\psi^+\rangle=(|00\rangle+|11\rangle+|22\rangle)/{\sqrt{3}}$
is a maximally entangled state. $\sigma_{\alpha}$ is just of the form (\ref{main state}),
a maximally entangled state $|\psi^+\rangle$
mixed with six pure separable states.
It is (PPT) separable for $2\leq \alpha \leq 3$,
PPT entangled for $3< \alpha \leq 4$, and NPT entangled for $4\leq
\alpha \leq 5$.

Utilizing the proof of Theorem 2, one can get a similar result for mixed states.

\begin{corollary}
For arbitrary mixed state $\rho_0$, if $\rho_0^{T_1}$ has $\frac{n(n-1)}{2}$ negative eigenvalues, then
\begin{eqnarray}
\rho=\lambda_0 \rho_0 +\sum_{i=1}^{K} \lambda_i |\chi_i\rangle\langle\chi_i|
\end{eqnarray}
is still NPT for $K\leq\frac{n(n-1)}{2}$, where $|\chi_i\rangle$ is an 
arbitrary pure separable state, $0<\lambda_i<1$, $i=0,1,\cdots, K$, $\sum_{i=0}^{K}\lambda_i=1$.
\end{corollary}

Our conclusions can be generalized to multipartite case. For
a multipartite quantum state $\rho$, we view it as a bipartite
quantum state with partition $\cal{Y}$ and  $\cal{\overline{Y}}$, where
the subsystems $\cal{Y}$ and subsystems $\cal{\overline{Y}}$ span the
whole quantum system, ${\cal{Y}}\cap {\cal{\overline{Y}}}=\emptyset$. Let $\rho^{T_{\cal{Y}}}$ denote the partial
transposition with respect to the subsystems $\cal{Y}$.
For a multipartite pure state $|\chi_0\rangle$, assume that
$(|\chi_0\rangle\langle\chi_0|)^{T_{\cal{Y}}}$ have $p_{\cal{Y}}$
negative eigenvalues. We set $p_{{\cal{Y}}_0}\equiv\max_{\cal{Y}} p_{\cal{Y}}$,
where the maximum goes over all possible partitions ${\cal{Y}}$.
Similar to the Theorem \ref{th1}, we have the following result,

\begin{theorem}
If $K\leq p_{{\cal{Y}}_0}-1$, then the quantum state $\rho=\lambda_0
|\chi_0\rangle\langle\chi_0| + \sum_{i=1}^{K} \lambda_i
|\chi_i\rangle\langle\chi_i|$ is still NPT, where $\{|\chi_i\rangle\}_{i=1}^K$
are arbitrary biseparable states under the partition between
${\cal{Y}}$ and $\cal{\overline{Y}}$, $0<\lambda_i<1$,
$i=0,1,\cdots, K$, $\sum_{i=0}^{K}\lambda_i=1$. Especially, $\rho$
is NPT if $\{|\chi_i\rangle\}_{i=1}^K$ are $K$ fully separable states.
\end{theorem}

We have studied the stability of the NPT property of an entangled pure state under
the mixture of arbitrary pure separable states. For bipartite pure state
with Schmidt number $n$ $(n>1)$, we have shown that it is still NPT under
mixing with no more than $\frac{n(n-1)}{2}-1$ arbitrary pure separable
states, with a generalization to multipartite cases.

For $n\geq 2$ and $K=\frac{n(n-1)}{2}$, we have the evidence that
the quantum state (\ref{main state}) is still NPT.
This result holds true at least for $n=2$, as shown in Theorem \ref{th n=2}.
However it still remains open whether the Theorem 2 is still valid
for $K\leq \frac{n(n-1)}{2}$.

\bigskip
{\noindent\bf Acknowledgement}
We would like to thank X. Li-Jost and T.G. Zhang for very useful and helpful
discussions.

\end{document}